\documentclass[11pt]{article}

\usepackage[utf8]{inputenc}
\usepackage[margin=1in]{geometry}
\usepackage{microtype}
\usepackage{amsmath,amssymb,amsthm}
\usepackage{booktabs,tabularx,array}
\usepackage{graphicx}
\usepackage{xcolor}
\usepackage{tikz}
\usetikzlibrary{arrows.meta,positioning,calc,fit,shapes.geometric}
\usepackage{pgfplots}
\pgfplotsset{compat=1.18}
\usepackage{algorithm}
\usepackage{algpseudocode}
\usepackage[font=small,labelfont=bf]{caption}
\usepackage{enumitem}
\usepackage[round,authoryear]{natbib}
\usepackage[colorlinks=true,linkcolor=navy,citecolor=navy,urlcolor=navy]{hyperref}

\definecolor{navy}{HTML}{1F4E79}
\definecolor{lightnavy}{HTML}{E8EEF5}
\definecolor{midgrey}{HTML}{9A9A9A}
\definecolor{darkgrey}{HTML}{555555}

\newtheorem{fact}{Result}

\newcolumntype{L}{>{\raggedright\arraybackslash}X}

\title{\textbf{Evolution of Market Microstructure in the Age of AI}\\[4pt]
\large From Double Auctions to Agentic Markets}
\author{Irene Aldridge\\
\small AbleBlox\\
\small \texttt{irene.aldridge@gmail.com}}
\date{September 2026}

\begin{document}
\maketitle

\begin{abstract}
\noindent Market microstructure studies how trading rules turn orders into prices and allocations. Those rules have been rebuilt repeatedly: for floor traders, electronic limit order books and high-frequency trading, batch auctions and dark pools, blockchains run by automated market makers and block builders, and now AI agents that discover, pay for and compete over resources. This survey traces that evolution through one question: \emph{can a market allocate scarce goods efficiently and fairly without participants revealing everything they know and want?} Each technological shift moved the binding constraint of market design, from trader rationality to speed, to control over transaction ordering, to the dimensionality of what participants can report. Impossibility results persist; only their cost moves. We review evidence on high-frequency trading, exchange and bilateral venues, automated market makers, extractable value and algorithmic collusion, and identify allocation as the missing layer of the agent-commerce protocol stack.


\end{abstract}

\tableofcontents
\newpage

% =====================================================================
\section{Introduction}
\label{sec:intro}
% =====================================================================

Every market must answer the same three questions: who trades, at what price, and in what order. Market microstructure theory, from the dealer models of \citet{glosten1985} and \citet{kyle1985} to the textbook synthesis of \citet{ohara1995}, studies how the answers depend on the rules of the trading venue and on what participants know. For most of that history the participants were people, and the rules were written for them. Three successive waves of automation have since changed who, or what, sits on each side of the book.

The first wave was electronic: continuous limit order books replaced the floor, and algorithms replaced the order clerk. The second was high-frequency: when execution moved to microseconds, speed itself became a scarce resource that firms competed for \citep{budish2015,aquilina2022}. The third, still underway, is \emph{agentic}: software agents built on large language models now search for, negotiate over and pay for goods and services with little or no human involvement \citep{allouah2025,zhu2025agents}. A parallel technological shift, the public blockchain, rebuilt trading infrastructure from scratch. It replaced dealers with algorithmic automated market makers (AMMs) and exchanges with block builders who decide which transactions are executed and in what order \citep{daian2020,xu2023sok}.

This survey argues that the lens that makes sense of all these changes is \emph{market design}: the explicit engineering of allocation rules subject to incentive, information and computational constraints. Viewed this way, the history of microstructure is a sequence of redesigns. Each was prompted by a technology that made the previous design's assumptions fail, and each was bounded by the same small set of impossibility theorems. The recurring question is whether scarce goods can be allocated efficiently and fairly without requiring participants to reveal everything they know and want.

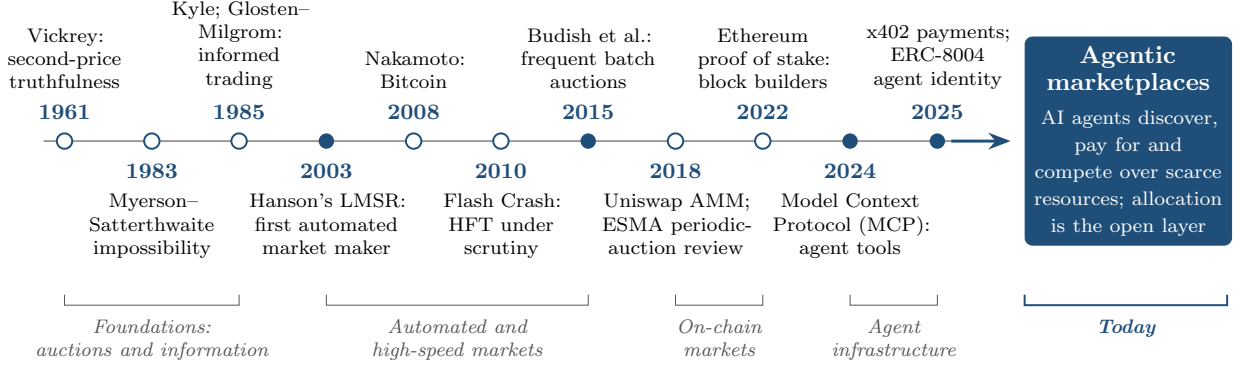
\begin{figure}[t]
\centering
\resizebox{\textwidth}{!}{%
\begin{tikzpicture}[x=1cm,y=1cm,font=\footnotesize]
  % axis running into the terminal stage
  \draw[thick,midgrey] (0,0) -- (13.75,0);
  \draw[-{Stealth[length=3.2mm]},very thick,navy] (13.0,0) -- (13.85,0);
  \foreach \x/\yr/\lab/\up/\hl in {
    0.3/1961/{Vickrey:\\second-price\\truthfulness}/1/0,
    1.55/1983/{Myerson--\\Satterthwaite\\impossibility}/0/0,
    2.8/1985/{Kyle; Glosten--\\Milgrom: informed\\trading}/1/0,
    4.05/2003/{Hanson's LMSR:\\first automated\\market maker}/0/1,
    5.3/2008/{Nakamoto:\\Bitcoin}/1/0,
    6.55/2010/{Flash Crash:\\HFT under\\scrutiny}/0/0,
    7.8/2015/{Budish et al.:\\frequent batch\\auctions}/1/1,
    9.05/2018/{Uniswap AMM;\\ESMA periodic-\\auction review}/0/0,
    10.3/2022/{Ethereum\\proof of stake:\\block builders}/1/0,
    11.55/2024/{Model Context\\Protocol (MCP):\\agent tools}/0/1,
    12.8/2025/{x402 payments;\\ERC-8004\\agent identity}/1/1}{
    \ifnum\hl=1
      \fill[navy] (\x,0) circle (0.11);
    \else
      \filldraw[fill=white,draw=navy,thick] (\x,0) circle (0.11);
    \fi
    \ifnum\up=1
      \node[navy,font=\footnotesize\bfseries] at (\x,0.42) {\yr};
      \node[align=center,text width=2.05cm,anchor=south,font=\scriptsize] at (\x,0.62) {\lab};
    \else
      \node[navy,font=\footnotesize\bfseries] at (\x,-0.42) {\yr};
      \node[align=center,text width=2.3cm,anchor=north,font=\scriptsize] at (\x,-0.62) {\lab};
    \fi
  }
  % terminal stage
  \node[draw=navy,fill=navy,rounded corners=3pt,text=white,align=center,
        minimum width=2.9cm,minimum height=2.2cm,text width=2.7cm,anchor=west] (end) at (14.05,0)
        {{\small\bfseries Agentic\\marketplaces}\\[3pt]
         {\scriptsize AI agents discover, pay for and compete over scarce resources; allocation is the open layer}};
  % era brackets
  \foreach \a/\b/\name in {0.3/2.8/{Foundations:\\auctions and information},
                           4.05/7.8/{Automated and\\high-speed markets},
                           9.05/10.3/{On-chain\\markets},
                           11.55/12.8/{Agent\\infrastructure}}{
    \draw[darkgrey,thin] (\a,-2.2) -- (\a,-2.35) -- (\b,-2.35) -- (\b,-2.2);
    \node[darkgrey,font=\scriptsize\itshape,anchor=north,align=center] at ({(\a+\b)/2},-2.42) {\name};
  }
  \draw[navy,thick] (14.05,-2.2) -- (14.05,-2.35) -- (16.95,-2.35) -- (16.95,-2.2);
  \node[navy,font=\scriptsize\bfseries\itshape,anchor=north] at (15.5,-2.42) {Today};
\end{tikzpicture}}
\caption{The evolution of market microstructure, from foundational auction theory through automated, high-speed and on-chain markets to agentic marketplaces. Filled markers denote designs whose ideas recur on-chain or in agentic markets later in the survey. The agent protocols of 2024--2025 standardize tools, payment and identity. The allocation layer of agentic marketplaces is the subject of Sections~\ref{sec:agents}--\ref{sec:matching}.}
\label{fig:timeline}
\end{figure}

\paragraph{The argument in brief.} Section~\ref{sec:foundations} recalls the foundational result: no bilateral trading mechanism can be efficient, incentive-compatible, individually rational and budget-balanced at once \citep{myerson1983}. Because this impossibility cannot be engineered away, every later design chooses which property to give up and where the loss shows up. We trace that choice through four eras.
\begin{enumerate}[leftmargin=*,itemsep=2pt]
  \item \textbf{Speed} (Section~\ref{sec:speed}). Continuous limit order books reward being first, creating a socially wasteful latency arms race. Frequent batch auctions change what ``simultaneous'' means and convert the speed race into price competition \citep{budish2015}. The price of that cure is timeliness: between clearing times no price can form, so periodic auctions prevent timely price discovery \citep{du2017,lee2026taiwan,esma2019periodic}.
  \item \textbf{Information} (Section~\ref{sec:information}). Once speed is neutralized, the information structure of the book becomes the lever. Recent queueing-theoretic work ranks welfare as dark pools $>$ lit exchanges $>$ batch auctions under moderate arrival rates \citep{aldridge2026mechanisms}. This refines rather than overturns the batch-auction case. At the opaque extreme sit bilateral dealer markets, where prices come from search and bargaining; electronic requests for quote, on-chain intents and agent-to-agent negotiation are their modern descendants (Section~\ref{sec:bilateral}).
  \item \textbf{Ordering control} (Sections~\ref{sec:revelation}--\ref{sec:onchain}). Blockchains reused Hanson's automated market maker \citep{hanson2003} and a batch architecture reminiscent of \citet{budish2015}, but gave control of each batch to a block builder. The result, maximal extractable value (MEV), is an institutionalized form of the front-running that batch auctions were designed to remove \citep{daian2020,qin2022}.
  \item \textbf{Dimensionality} (Sections~\ref{sec:agents}--\ref{sec:matching}). AI agents now have protocols for tools \citep{anthropic2024mcp}, payment \citep{coinbase2025x402} and identity \citep{erc8004}, but none of these decides who gets a scarce resource. Agents can cheaply report a few feature weights but not full rankings, so a new allocation problem appears. Spectral feature matching \citep{aldridge2026matching} is one answer: its guarantee is conditional, and its failures can be checked in advance.
\end{enumerate}
Section~\ref{sec:synthesis} draws these threads together and Section~\ref{sec:conclusion} concludes.

\paragraph{Scope and positioning.} Excellent surveys exist on each strand: classical microstructure \citep{ohara1995,ohara2015}, high-frequency trading \citep{aldridge2013hft}, decentralized exchanges \citep{xu2023sok}, and matching theory. Our contribution is to read them as one sequence of design problems with a common structure, and to carry that structure forward into markets populated by AI agents. The survey draws on three companion pieces by the author: an analysis of lit, dark and batch mechanisms \citep{aldridge2026mechanisms}, a literature review of crypto microstructure prepared for the Federal Reserve Bank of New York \citep{aldridge2025crypto}, and a working paper on multi-dimensional matching \citep{aldridge2026matching}. Where we report their results we say so explicitly.

% =====================================================================
\section{Foundations: What a Trading Mechanism Can and Cannot Do}
\label{sec:foundations}
% =====================================================================

\subsection{The double auction and the clearing price}

The canonical trading mechanism is the double auction. Buyers submit bids, sellers submit asks, and the mechanism sets a price at which trade clears. Sorting bids in descending order traces out a demand schedule; sorting asks in ascending order traces out a supply schedule; a uniform clearing price sits where the two cross. Every buyer whose bid is above the price and every seller whose ask is below it trades.

The design question is whether traders will submit their true values. For one-sided auctions \citet{vickrey1961} gave a celebrated answer: if the winner pays the \emph{second}-highest bid, then the price a bidder pays never depends on her own bid, so shading can only lose profitable trades and overbidding can only win unprofitable ones. Truthful bidding is a weakly dominant strategy in every state of the world.

\subsection{The two-sided impossibility}

Two-sided markets are harder. With private values on both sides, each trader's report affects the price she faces, and incentives to shade appear on both sides of the book. \citet{chatterjee1983} showed that in the bilateral case strategic bidding leaves a band of efficient trades near the diagonal of the buyer-value/seller-cost plane unconsummated. \citet{myerson1983} proved that this is not a failure of any particular design but a mathematical impossibility.

\begin{fact}[\citealp{myerson1983}]
When a buyer's and a seller's privately known valuations are independently distributed with overlapping supports, no bilateral trading mechanism is simultaneously ex-post efficient, Bayesian incentive-compatible, interim individually rational and budget-balanced without an outside subsidy.
\end{fact}

The theorem forces a choice, summarized in Table~\ref{tab:impossibility}. \citet{mcafee1992} gives the best-known escape: compute the efficient number of trades, then drop the marginal buyer--seller pair if needed so that prices can be set from the excluded bids. The mechanism is dominant-strategy truthful and budget-balanced; the cost is the loss of at most one near-indifferent trade. Vickrey--Clarke--Groves mechanisms keep efficiency and truthfulness but run a deficit that someone must subsidize.

\begin{table}[t]
\centering
\caption{The Myerson--Satterthwaite impossibility forces a choice of which property to give up.}
\label{tab:impossibility}
\begin{tabularx}{\textwidth}{@{}p{3.3cm}LL@{}}
\toprule
\textbf{Give up} & \textbf{What remains} & \textbf{Example} \\
\midrule
Exact efficiency & Budget balance, truthfulness, participation & \citet{mcafee1992} trade reduction \\
Budget balance & Efficiency, truthfulness, participation & VCG with an external subsidizer \\
Individual rationality & Efficiency, budget balance, truthfulness & Forced-participation schemes \\
Truthfulness & Efficiency, budget balance, participation & Posted-price bargaining \\
\bottomrule
\end{tabularx}
\end{table}

\subsection{Why real double auctions work anyway}

If the theory is so pessimistic, why do continuous double auctions work well in practice? Laboratory evidence answered early. \citet{smith1962} found that continuous double auctions reach near-full efficiency almost immediately, even with few traders and no knowledge of others' values. \citet{gode1993} went further: ``zero-intelligence'' traders who bid randomly, subject only to a budget constraint that forbids loss-making trades, achieve allocative efficiency close to that of human traders. Unconstrained random traders do not.

The lesson has shaped market design ever since: \emph{market structure, not trader sophistication, does most of the work}. It also foreshadows the agentic era. If the rules of a well-designed market can extract efficiency from nearly random traders, the rules matter even more when the traders are software whose objectives and strategies may be opaque (Section~\ref{sec:agents}).

\subsection{Information and the market maker}

The classical microstructure models add the other essential ingredient: asymmetric information. In \citet{glosten1985} a competitive dealer sets bid and ask quotes knowing that some counterparties are better informed; the spread compensates for adverse selection. In \citet{kyle1985} an informed trader camouflages orders among noise traders and a market maker sets prices from aggregate order flow. Both frameworks identify the \emph{liquidity provider} as the agent who bears adverse selection and the \emph{information structure} of the market as the determinant of prices. The same two elements reappear, in algorithmic form, in every design that follows: the high-frequency market maker, the dark pool, the AMM's liquidity pool and the agent marketplace.

% =====================================================================
\section{The Speed Era: Continuous Trading and the Arms Race}
\label{sec:speed}
% =====================================================================

\subsection{Algorithms become the market makers}

Electronic continuous limit order books process each order the instant it arrives, under price-time priority. By the late 2000s, market making on most venues had been taken over by high-frequency trading (HFT) firms that post limit orders and earn the spread \citep{menkveld2013}. The early evidence on liquidity was largely favorable. Algorithmic trading narrowed spreads and made quotes more informative \citep{hendershott2011}. HFTs trade in the direction of permanent price changes and against transitory pricing errors \citep{brogaard2014}. \citet{ohara2015} argued that high-frequency markets require rethinking microstructure theory itself, since the trade is no longer the natural unit of analysis. The May 2010 Flash Crash showed the fragility side of the same ecosystem: HFTs did not trigger the crash but amplified its dynamics as liquidity evaporated \citep{kirilenko2017}.

\subsection{The latency arms race}

\citet{budish2015} identified a structural flaw in continuous trading itself. When a public signal moves the value of an asset, every liquidity provider's quote becomes stale at once, and the first trader to reach the exchange can ``snipe'' it. Because correlations between closely related instruments that hold at human timescales break down at microsecond horizons, such opportunities recur mechanically, all day, every day. The resulting race has the structure of a prisoner's dilemma (Table~\ref{tab:pd}): each firm must invest in speed to avoid being picked off, and in equilibrium all firms pay the fixed cost while the arbitrage rents are competed back to the spread paid by investors.

\begin{table}[t]
\centering
\caption{The latency race as a symmetric prisoner's dilemma \citep{budish2015}.}
\label{tab:pd}
\begin{tabular}{@{}lcc@{}}
\toprule
 & \textbf{Rival stays slow} & \textbf{Rival invests in speed} \\
\midrule
\textbf{You stay slow} & Both keep profits & You are picked off \\
\textbf{You invest in speed} & You pick off the rival & Both pay the cost; rents competed away \\
\bottomrule
\end{tabular}
\end{table}

\citet{aquilina2022} measured these races directly using message-level data from the London Stock Exchange, including failed orders that are invisible in ordinary trade data. They find that races are frequent, last microseconds, and account for a substantial share of trading volume. The implied ``latency arbitrage tax'' on liquidity is economically meaningful in aggregate, which confirms that the arms race is a first-order feature of continuous markets rather than a theoretical curiosity.

\subsection{Frequent batch auctions}

The proposed remedy discretizes time. Under frequent batch auctions, orders arriving within a short interval are treated as simultaneous and cleared together in a uniform-price sealed-bid double auction \citep{budish2014,budish2015}. Within a batch, price-time priority is replaced by pro-rata allocation at the clearing price, so arriving a microsecond earlier confers no advantage. Discreteness converts competition on speed into competition on price. Because liquidity providers no longer need to price in the risk of being sniped, the model predicts narrower spreads.

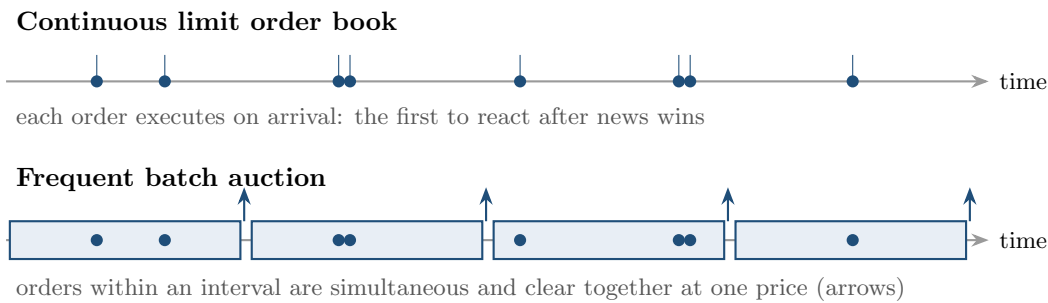
\begin{figure}[t]
\centering
\begin{tikzpicture}[font=\footnotesize]
  % continuous
  \node[anchor=west,font=\small\bfseries] at (0,2.2) {Continuous limit order book};
  \draw[-{Stealth},thick,midgrey] (0,1.4) -- (13,1.4) node[right,black]{time};
  \foreach \x/\l in {1.2/1,2.1/2,4.4/3,4.55/4,6.8/5,8.9/6,9.05/7,11.2/8}{
    \fill[navy] (\x,1.4) circle (0.08);
    \draw[navy] (\x,1.4) -- (\x,1.75);
  }
  \node[align=left,anchor=west,darkgrey] at (0,0.95) {each order executes on arrival: the first to react after news wins};
  % batch
  \node[anchor=west,font=\small\bfseries] at (0,0.1) {Frequent batch auction};
  \draw[-{Stealth},thick,midgrey] (0,-0.7) -- (13,-0.7) node[right,black]{time};
  \foreach \s in {0,3.2,6.4,9.6}{
    \draw[navy,thick,fill=lightnavy] (\s+0.05,-0.95) rectangle (\s+3.1,-0.45);
  }
  \foreach \x in {1.2,2.1,4.4,4.55,6.8,8.9,9.05,11.2}{
    \fill[navy] (\x,-0.7) circle (0.08);
  }
  \foreach \s/\n in {3.15/1,6.35/2,9.55/3,12.75/4}{
    \draw[-{Stealth},navy,thick] (\s,-0.45) -- (\s,0.0);
  }
  \node[align=left,anchor=west,darkgrey] at (0,-1.35) {orders within an interval are simultaneous and clear together at one price (arrows)};
\end{tikzpicture}
\caption{Continuous trading versus frequent batch auctions. Batching removes the value of arriving microseconds earlier within an interval.}
\label{fig:batch}
\end{figure}

\subsection{Batch mechanisms in practice}

Periodic auctions are not only a theoretical proposal. In Europe, periodic auction books run alongside continuous lit trading on several venues (for example, the Cboe Europe periodic auctions book and Deutsche B\"orse's Xetra auction mechanisms). Their growth after the MiFID~II dark-trading caps prompted a formal review by the European Securities and Markets Authority \citep{esma2018periodic,esma2019periodic}, whose findings on price formation we discuss below. In the United States, IEX adopted a different, non-batch response to the same race: a uniform 350-microsecond delay on inbound orders. The batch interval is itself a design parameter with costs on both sides. Longer intervals allow prices to go stale, while shorter ones leave residual sniping risk, so total cost is U-shaped in the interval length.

\subsection{The cost of batching: delayed price discovery}
\label{sec:batch-discovery}

Batching buys protection from sniping by giving up timeliness. In a continuous book every order that crosses the spread is a trade, and every trade is a price that others can observe and act on. In a periodic auction, no trade and hence no transaction price can form between clearing times. Information that arrives just after an auction can reach prices only at the next one, while quotes and resting interest are, by design, stale. \emph{Periodic auctions therefore prevent timely price discovery}: the question is only how large the delay is and who bears it. Theory, a natural experiment in exchange redesign and regulatory evidence all point to the same cost.

\paragraph{Theory.} \citet{madhavan1992} compares continuous and periodic call markets. Pooling orders in a call market can sustain trading under information asymmetry severe enough to shut down a continuous market, but prices are revealed only at the discrete clearing times. \citet{du2017} solve for the trading frequency that maximizes allocative efficiency in a model of sequential double auctions. When news arrives at scheduled times, the optimal frequency matches the frequency of information arrival. When news arrives stochastically, as most market-moving information does, the optimal frequency \emph{far exceeds} the information arrival rate, because infrequent auctions leave traders unable to respond promptly to new information. Section~\ref{sec:information} adds a welfare channel: in \citet{aldridge2026mechanisms}, batching forces even the most urgent traders to wait $T/2$ on average, and this is one of the two reasons batch auctions rank below lit exchanges.

\paragraph{Evidence from a market redesign.} The Taiwan Stock Exchange offers a rare natural experiment. \citet{twu2018} study the shortening of Taiwan's call-auction interval from 25 to 5 seconds between 2010 and 2014, and find mixed effects on pricing-efficiency measures but better overall market quality at the shorter interval. \citet{lee2026taiwan} exploit Taiwan's 2020 switch from frequent batch auctions to continuous trading. Liquidity and price efficiency improved significantly for mid- and small-capitalization stocks, which also earned positive abnormal returns, whereas the effect on large-capitalization efficiency was marginal. Faster traders captured more of the gains, and individual investors in smaller stocks lost more. Batching thus slows price discovery most where information is scarcest and matters most: in less liquid stocks.

\paragraph{Regulatory evidence.} ESMA's review of European periodic auctions found that many were not genuinely price-forming: 99.9\% of orders submitted to them were pegged to reference prices set elsewhere, and about 80\% crossed at the midpoint of the continuous market's spread \citep{esma2019periodic}. ESMA proposed treating mid-point-only systems as non-price-forming. Batch venues that import their prices from continuous books do not perform price discovery; they consume it.

The design implication is a genuine trade-off, not a verdict against batching. Shorter intervals reduce staleness but restore part of the speed race, and batching may remain valuable at the open and close or in thin instruments. As a general replacement for continuous trading, however, periodic auctions exchange one cost, the latency arms race, for another, delayed price discovery.

The broader lesson of the speed era is that market design can neutralize an arms race without banning speed or restricting who may trade. It does so by \emph{redefining what counts as simultaneous}. The same lever, control over the sequencing of orders, reappears with very different consequences on blockchains (Section~\ref{sec:onchain}).

% =====================================================================
\section{The Information Era: Lit, Dark and Batch Venues}
\label{sec:information}
% =====================================================================

\subsection{Three venues as one queueing system}

Eliminating the speed race does not by itself establish that batch auctions maximize welfare. Modern equity markets offer at least three architectures: lit exchanges, dark pools and batch auctions. \citet{aldridge2026mechanisms} models all three as the same queueing system with one element changed each time: the information traders have about the book and the discipline by which orders are served (Table~\ref{tab:venues}).

\begin{table}[t]
\centering
\caption{Three market architectures as variants of one queueing system \citep{aldridge2026mechanisms}.}
\label{tab:venues}
\begin{tabularx}{\textwidth}{@{}lLLL@{}}
\toprule
\textbf{Architecture} & \textbf{Order book} & \textbf{Service discipline} & \textbf{Market-order cutoff depends on book?} \\
\midrule
Lit exchange & Publicly observable & First-come-first-served & Yes: $C^*(V,B)$ \\
Dark pool & Hidden until trade reported & First-come-first-served & No: constant $\bar C^*$ \\
Batch auction & Accumulates over interval $T$ & Service in random order & No: $\tau^*(V,C)$ \\
\bottomrule
\end{tabularx}
\end{table}

Two foundational queueing results point in opposite directions. \citet{leshno2022} shows that serving agents in random order can outperform first-come-first-served because it smooths the mismatch caused by fluctuations in waiting times. That argument favors batch auctions over lit books. \citet{che2025queue} show that an uninformed first-come-first-served queue, which reveals nothing beyond recommendations, is optimal because hiding queue position removes the incentive to exit. That argument favors dark pools. Neither model was built for financial markets, which have an escape valve absent from classical queues: an impatient trader can bypass the queue entirely by paying the spread with a market order.

\subsection{Model and welfare ranking}

In \citet{aldridge2026mechanisms} each trader $i$ with valuation $V_i$, side $s_i\in\{-1,+1\}$ and waiting cost $C_i$ trades off execution price, waiting and a fixed cost $K$:
\begin{equation}
U_i(p,w_i) \;=\; s_i\,(V_i - p) \;-\; C_i\, w_i \;-\; K .
\end{equation}
The specification is identical across mechanisms; only the information structure and service discipline differ. Only on the lit exchange do traders condition their strategy on the visible book state $B$. The paper's main result ranks welfare.

\begin{fact}[\citealp{aldridge2026mechanisms}, Theorem 2]
For arrival rates $\lambda\in[\lambda_{\min},\lambda_{\max}]$, bounded adverse selection $\Delta<\Delta_{\max}$ and a dark-pool reporting delay $\delta<T/2$,
\[
W_{\text{DARK}} \;>\; W_{\text{LIT}} \;>\; W_{\text{BATCH}} .
\]
\end{fact}

Two distinct inefficiencies drive the ranking, and execution prices cancel from both because they are transfers between buyer and seller. \emph{Dark beats lit} because observing the book induces costly timing games in which traders delay or rush to improve queue position; opacity removes the incentive, while urgent traders can still execute instantly with market orders. \emph{Lit beats batch} because batching forces even the most urgent traders to wait $T/2$ on average, and pro-rata rationing pushes marginal traders out of the market altogether.

\begin{figure}[t]
\centering
\begin{tikzpicture}
\begin{axis}[
  ybar, bar width=26pt, width=0.62\textwidth, height=5.6cm,
  ymin=0, ymax=0.34, ylabel={Aggregate welfare},
  symbolic x coords={Dark pool,Lit exchange,Batch auction},
  xtick=data, nodes near coords, nodes near coords align={vertical},
  every node near coord/.append style={font=\footnotesize,/pgf/number format/fixed,/pgf/number format/fixed zerofill,/pgf/number format/precision=4},
  axis x line*=bottom, axis y line*=left, enlarge x limits=0.25,
  tick label style={font=\footnotesize}, label style={font=\footnotesize},
  ymajorgrids, grid style={gray!20}]
\addplot[fill=navy,draw=navy] coordinates {(Dark pool,0.2819) (Lit exchange,0.2410) (Batch auction,0.1589)};
\end{axis}
\end{tikzpicture}
\caption{Simulated aggregate welfare at baseline calibration ($n=20{,}000$ traders, $\lambda=5$, $\Delta=2$, $T=1$). Participation rates are 46.1\%, 43.6\% and 39.9\%. The dark pool exceeds the lit exchange by 17\% and the lit exchange exceeds batch by 52\%. Source: \citet{aldridge2026mechanisms}, Table~3.}
\label{fig:welfare}
\end{figure}
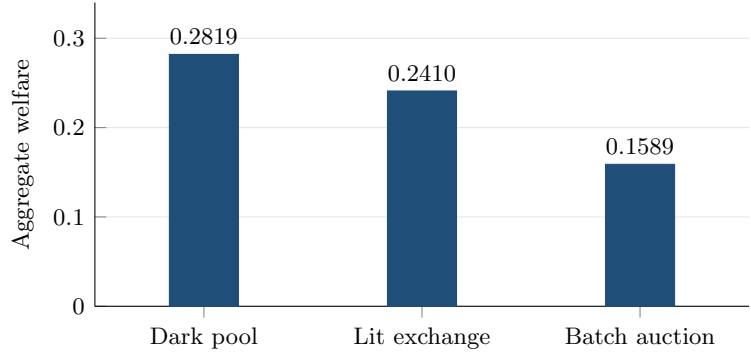

Figure~\ref{fig:welfare} reports the calibrated magnitudes. Over a $20\times20$ grid of arrival rates and valuation dispersions, the full ranking holds almost everywhere. It fails only in very thin markets ($\lambda\lesssim 2.5$), where just $W_{\text{DARK}}>W_{\text{LIT}}$ survives, and the failure region shrinks as dispersion grows. The model also identifies the real cost of opacity. It is not that uninformed traders receive worse prices, which is only a transfer. It is that marginal traders anticipate poor pricing and exit, and the surplus from their trades is lost. This \emph{participation channel} connects the result to the long-standing debate on whether dark pools harm price discovery \citep{zhu2014}.

\subsection{Regulatory relevance}

Regulators on both sides of the Atlantic have recently revisited these architectures. In December 2022 the U.S.\ Securities and Exchange Commission proposed four equity-market-structure rules, including an Order Competition Rule that would have required certain segmented retail orders to be exposed to auctions before internalization, and a Regulation Best Execution \citep{sec2022proposals}. Both of those proposals were withdrawn in June 2025 \citep{sec2025withdrawal}. In April 2026 ESMA launched a call for evidence on the structure of European equity markets covering lit continuous trading, closing and frequent batch auctions, systematic internalisers, dark trading and post-trade transparency \citep{esma2026structure}. The welfare decomposition above has a direct implication for such reviews: shortening batch intervals helps batch auctions, but transparency mandates do not, because opacity was never the batch auction's problem. Mandating pre-trade disclosure in dark venues would remove the very feature that generates their welfare advantage. This refines rather than overturns Section~\ref{sec:speed}. Batch auctions still eliminate the speed race, but on this welfare measure they are not the ceiling.

% =====================================================================
\section{Bilateral Markets: Dealers, Search and Requests for Quote}
\label{sec:bilateral}
% =====================================================================

\subsection{Why bilateral trading persists}

Exchanges are the exception, not the rule. Most corporate and municipal bonds, interest-rate and credit swaps, foreign exchange and many large crypto trades change hands \emph{bilaterally}: a customer negotiates with a dealer, one counterparty at a time, over the counter (OTC). \citet{bessembinder2020survey} survey the microstructure of these fixed-income markets. Bilateral trade is also the literal setting of the Myerson--Satterthwaite theorem (Section~\ref{sec:foundations}): two parties with private valuations bargain over a single trade, so some efficient trades must fail. The canonical bargaining model is the alternating-offers game of \citet{rubinstein1982}, in which patience determines how the surplus is split. Bilateral markets sit at the opaque extreme of the venue spectrum in Section~\ref{sec:information}: there is no public book, quotes are private, and each trader sees only the prices she is offered.

\subsection{Search, bargaining and dealer networks}

\citet{duffie2005otc} model OTC markets as search markets. Investors must find a counterparty before they can trade, and prices are set by bargaining. Bid--ask spreads therefore reflect search frictions and the relative bargaining power of investors and dealers, not only inventory and adverse selection. Spreads narrow as investors find counterparties more easily or can meet more dealers \citep[see also][]{duffie2012dark}. \citet{zhu2012otc} shows how opacity shapes this search. A seller who contacts dealers sequentially reveals information through repeated contact, and dealers who infer that the seller has been turned down elsewhere quote worse prices. Real OTC markets are organized as networks. \citet{li2019dealer} document a core--periphery structure of dealers in which trading costs and speed of execution depend on a dealer's position in the network.

Empirically, bargaining power in opaque bilateral markets favors the dealer, especially against small customers. In municipal bonds, \citet{green2007muni} find that dealers earn substantial markups that decline sharply with trade size, so retail investors pay the most. Post-trade transparency changes the balance. When the TRACE system began publishing corporate bond trade prices, execution costs fell for the newly transparent bonds \citep{bessembinder2006trace,edwards2007}. In bilateral markets, publishing prices after the fact performs part of the price-discovery work that a public order book performs before the trade.

\subsection{Electronic requests for quote}

Electronic trading has entered bilateral markets through the request for quote (RFQ). A customer sends a trade request to several dealers at once, and the best response wins. \citet{hendershott2015click} show that these one-sided electronic auctions are a viable source of liquidity even in inactively traded bonds. They describe the RFQ as a natural compromise between bilateral search and the continuous double auction of an exchange. Using platform and regulatory data, \citet{ohara2021bond} find that electronic RFQ trading, though still limited in scope, has substantial effects on transaction costs and execution quality across electronic, voice and inter-dealer trading. Regulation has pushed in the same direction: the Dodd--Frank Act required many swaps to trade on swap execution facilities. Yet \citet{riggs2020sef} find that customers of index credit default swaps typically contact only a few dealers. The benefit of wider competition is limited by a winner's curse, since dealers who win a widely shopped request fear they have mispriced, and by customer--dealer relationships.

The RFQ is itself a periodic, sealed-bid auction triggered by one trader. It inherits both sides of Section~\ref{sec:batch-discovery}. It concentrates competition at a moment chosen by the customer, but it produces no continuous public price, and shopping a request widely leaks information about the customer's intentions.

\subsection{Bilateral trading on-chain and between agents}

Bilateral mechanisms are returning in decentralized finance. Instead of trading against an AMM pool, a user can sign an \emph{intent}, a statement of what she wants to trade and at what worst price, and let competing ``solvers'' or market makers bid to fill it. These order-flow auctions are on-chain RFQs. \citet{bachu2024ofa} measure the price improvement they deliver relative to AMM execution on two major platforms, finding average gains of about 4--5 basis points that come mainly from better access to liquidity for larger trades. \citet{canidio2024intents} design fair combinatorial auctions for batches of such intents. Filling an intent privately also keeps it out of the public mempool, which is a defense against the extractable value of Section~\ref{sec:onchain}.

Agentic markets are bilateral by default. When a user's agent negotiates directly with a seller's agent, the interaction is bilateral bargaining under private information, and the Myerson--Satterthwaite logic applies: some efficient trades will fail, and the split of surplus depends on bargaining power. \citet{zhu2025agents} find that such unmediated agent-to-agent negotiations can produce systematically imbalanced outcomes. The x402 protocol of Section~\ref{sec:agents} implements the simplest bilateral mechanism of all, a posted price: the server names a price and the client pays or walks away. Posted prices are simple and give the buyer no reason to misreport, but they lose every trade in which the buyer's value lies below the price and the seller's cost below the buyer's value.

\begin{table}[t]
\centering
\caption{Bilateral trading mechanisms, from voice dealers to agents.}
\label{tab:bilateral}
\begin{tabularx}{\textwidth}{@{}>{\raggedright\arraybackslash}p{3.2cm}LL@{}}
\toprule
\textbf{Mechanism} & \textbf{How a price is formed} & \textbf{Main friction} \\
\midrule
Voice OTC dealer market & Sequential search and bilateral bargaining \citep{duffie2005otc} & Search costs; dealer bargaining power; opacity \citep{green2007muni} \\
Electronic RFQ & One-sided sealed-bid auction among invited dealers \citep{hendershott2015click} & Few dealers contacted; winner's curse; information leakage \citep{riggs2020sef} \\
Post-trade transparency (TRACE) & Bilateral quotes, publicly reported afterwards & Price discovery after the fact \citep{bessembinder2006trace,edwards2007} \\
On-chain intents / order-flow auctions & Solvers compete to fill a signed intent \citep{bachu2024ofa} & Concentration among solvers; fairness across intents \citep{canidio2024intents} \\
Agent-to-agent negotiation & Bargaining between software agents & Imbalanced outcomes without a mediator \citep{zhu2025agents} \\
Posted price (x402) & Seller names a price; buyer accepts or leaves & Lost trades below the posted price \\
\bottomrule
\end{tabularx}
\end{table}

Table~\ref{tab:bilateral} summarizes. Across all these settings, the evolution of bilateral markets repeats the pattern of the survey: technology adds competition (RFQ, intents), transparency (TRACE) or automation (agents) to one-to-one trading, but the Myerson--Satterthwaite bound means that some centralized mechanism is needed to recover the efficient trades that bilateral bargaining leaves on the table.

% =====================================================================
\section{Designing for What Participants Report: Revelation and Automated Market Makers}
\label{sec:revelation}
% =====================================================================

\subsection{The revelation principle and its limits}

Designing a mechanism appears to require searching an unbounded space of games: ascending auctions, sealed bids, multi-round negotiations. The revelation principle collapses the search. Any equilibrium outcome of any mechanism can be reproduced by a direct mechanism in which agents truthfully report their types and the mechanism plays their equilibrium strategies for them \citep{myerson1979,myerson1981}. The simplification has a cost that becomes central in the agentic era. ``Truthful'' means truthful about \emph{what the mechanism asks for}, and the principle does not guarantee that the truthful direct mechanism is simple, fast or computable.

\citet{gibbard1973} and \citet{satterthwaite1975} then showed that with three or more outcomes and unrestricted preferences, the only strategy-proof social choice rules are dictatorial. Escaping the theorem requires restricting one of its conditions (Table~\ref{tab:gs}). Each escape route corresponds to a family of practical mechanisms, and the one relevant for AI agents, restricting the preference domain, is developed in Section~\ref{sec:matching}.

\begin{table}[t]
\centering
\caption{Escape routes from the Gibbard--Satterthwaite theorem.}
\label{tab:gs}
\begin{tabularx}{\textwidth}{@{}p{4cm}L@{}}
\toprule
\textbf{Restrict} & \textbf{Example} \\
\midrule
Outcomes to two & Majority rule is strategy-proof between two options \\
The preference domain & Single-peaked preferences make the median voter rule strategy-proof; low-dimensional feature preferences (Section~\ref{sec:matching}) \\
Allow money & Vickrey--Clarke--Groves mechanisms, the route taken by auctions \\
\bottomrule
\end{tabularx}
\end{table}

\subsection{Prediction markets and the first automated market maker}

Prediction markets restrict the domain to a single uncertain event, and good mechanisms become possible. A contract's price is a forecast: the level at which no one wants to trade further. \citet{wolfers2004} show that such prices forecast outcomes about as well as expert models.

\paragraph{Prediction markets run on limit order books.} In practice, the prediction markets that carry real volume are organized as limit order books, the double auction of Section~\ref{sec:foundations} in continuous form. The Iowa Electronic Markets operate a continuous double auction with market and limit orders \citep{berg2008}. Intrade was an exchange on which traders posted their own bids and asks, and \citet{rothschild2016} use its transaction-level data to study trading strategies and market microstructure. The two largest venues today follow the same design (Table~\ref{tab:pm}). Kalshi, designated by the CFTC as a contract market in November 2020 \citep{cftc2020kalshi}, displays resting bids and asks for each event contract \citep{kalshi2026orderbook}. Polymarket began with an automated market maker and moved to a hybrid central limit order book, in which an operator matches signed orders off-chain and trades settle on-chain \citep{tsang2026polymarket}.

\begin{table}[t]
\centering
\caption{Trading mechanisms of major prediction markets.}
\label{tab:pm}
\begin{tabularx}{\textwidth}{@{}>{\raggedright\arraybackslash}p{3.3cm}LL@{}}
\toprule
\textbf{Venue} & \textbf{Mechanism} & \textbf{Source} \\
\midrule
Iowa Electronic Markets & Continuous double auction; market and limit orders & \citet{berg2008} \\
Intrade & Exchange with trader-posted bids and asks & \citet{rothschild2016} \\
Kalshi & CFTC-designated contract market; order book of resting bids and asks & \citet{cftc2020kalshi,kalshi2026orderbook} \\
Polymarket & Initially an AMM; now a hybrid CLOB with off-chain matching and on-chain settlement & \citet{tsang2026polymarket} \\
Hanson-style markets & Logarithmic market scoring rule (automated market maker) & \citet{hanson2003,hanson2007} \\
\bottomrule
\end{tabularx}
\end{table}

The order book's weakness is thin liquidity. A thinly traded contract has few resting orders, so incoming traders often find no counterparty.

\citet{hanson2003,hanson2007} addressed thin liquidity by replacing the matched counterparty with an algorithm. Under the logarithmic market scoring rule (LMSR), every trader transacts against a market maker whose prices derive from a smooth convex cost function over the vector $q$ of outstanding shares in each of $n$ outcomes:
\begin{equation}
C(q) = b\,\log\!\Big(\sum_{i=1}^{n} e^{q_i/b}\Big), \qquad
p_i(q) = \frac{\partial C}{\partial q_i} = \frac{e^{q_i/b}}{\sum_j e^{q_j/b}} .
\label{eq:lmsr}
\end{equation}
Buying $\Delta q$ costs $C(q+\Delta q)-C(q)$. Prices are always quoted and sum to one, and the market maker's worst-case loss is bounded by $b\log n$ regardless of volume. The liquidity parameter $b$ is a designed subsidy: larger $b$ buys deeper liquidity at a larger, but known, maximum cost.

The LMSR is the conceptual ancestor of the constant-function market makers that dominate decentralized finance. A cost function replaces the counterparty; liquidity is always available, 24 hours a day, without human supervision; and the designer's problem becomes choosing the shape of the curve.

The two market types have migrated in opposite directions. Decentralized spot exchanges moved from order books to AMMs (Section~\ref{sec:onchain}), while prediction markets, once volume arrived, moved from AMMs to order books, as Polymarket's redesign illustrates. The pattern is consistent with the division of labor suggested by the LMSR's bounded subsidy. An algorithmic market maker is worth its cost when counterparties are scarce, but once enough traders arrive, a limit order book lets them quote their own prices and removes the need for a subsidized counterparty.

% =====================================================================
\section{On-Chain Microstructure: Blockchains, AMMs and Extractable Value}
\label{sec:onchain}
% =====================================================================

\subsection{What institutional adoption requires}

A blockchain is a replicated database in which each block carries a cryptographic summary of its predecessor, so altering history requires recomputing the entire chain \citep{haber1991,nakamoto2008}. Its promise for finance is 24/7 automated settlement through smart contracts \citep{cong2019}. \citet{aldridge2025crypto} distills institutional requirements into six market-quality bars: (i) fast and certain settlement; (ii) institutional-grade throughput; (iii) predictable transaction (``gas'') costs; (iv) minimal extractable value, fraud and attack surface; (v) efficient matching of buyers and sellers; and (vi) privacy of business terms. Bars (iv) and (v) are market-design problems and occupy most of this section.

\begin{table}[t]
\centering
\caption{Indicative maximum throughput, transactions per second. Figures as compiled in \citet{aldridge2025crypto} from Visa, NYSE, \emph{Forbes} (2009) and public blockchain sources. Capacity definitions differ across venues, so comparisons are order-of-magnitude only.}
\label{tab:tps}
\begin{tabular}{@{}lr@{}}
\toprule
\textbf{System} & \textbf{Max.\ transactions per second} \\
\midrule
NASDAQ & 250{,}000 \\
NYSE & 86{,}000 \\
Visa & 65{,}000 \\
Ethereum (base layer) & 15--30 (up to $\sim$1{,}000 with Layer-2 rollups) \\
Bitcoin & 3--7 \\
\bottomrule
\end{tabular}
\end{table}

Public blockchains fall short on the first two bars (Table~\ref{tab:tps}). Throughput is four to five orders of magnitude below traditional venues. Layer-2 ``rollups'' batch transactions off-chain and post results back. Optimistic rollups assume validity unless challenged with a fraud proof during a withdrawal window of about seven days. Zero-knowledge rollups prove validity cryptographically at higher computational cost. Settlement finality varies from probabilistic and never absolute on Bitcoin, to economic finality after roughly a quarter of an hour on proof-of-stake Ethereum, to deterministic sub-second finality on newer chains designed for payments such as Circle's Arc, which also offers opt-in shielded balances \citep{circle2025arc}. Traditional securities settle on a T+1 cycle but with legal certainty. Designers also face the ``scalability trilemma'' between throughput, security and decentralization. Buterin's heuristic for decentralization is whether each half of a system, cut in two, could continue operating independently \citep{buterin2017decentralization}. One practical compromise is quasi-decentralization: geographically distributed nodes run by a small set of vetted operators.

\subsection{Constant-function market makers}

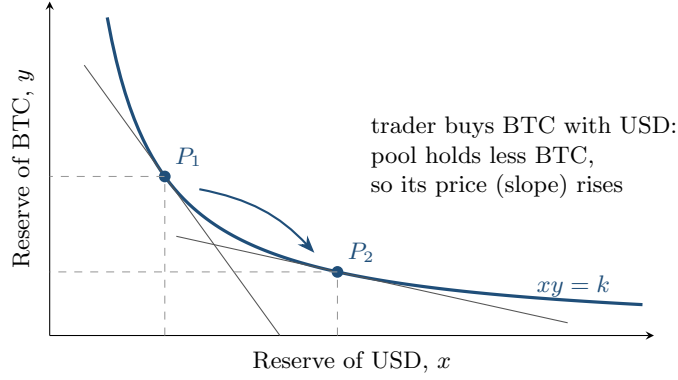
\begin{figure}[t]
\centering
\begin{tikzpicture}
\begin{axis}[width=0.58\textwidth,height=6cm,xmin=0,xmax=10.5,ymin=0,ymax=10.5,
  axis lines=left, xlabel={Reserve of USD, $x$}, ylabel={Reserve of BTC, $y$},
  xtick=\empty, ytick=\empty, label style={font=\footnotesize}, clip=false]
  \addplot[navy,very thick,domain=1.0:10.3,samples=120]{10/x};
  % P1 at x=2 (y=5), slope -2.5 ; P2 at x=5 (y=2), slope -0.4
  \addplot[darkgrey,thin,domain=0.6:4.0]{5-2.5*(x-2)};
  \addplot[darkgrey,thin,domain=2.2:9.0]{2-0.4*(x-5)};
  \fill[navy] (axis cs:2,5) circle (2.2pt) node[above right,font=\footnotesize]{$P_1$};
  \fill[navy] (axis cs:5,2) circle (2.2pt) node[above right,font=\footnotesize]{$P_2$};
  \draw[dashed,midgrey] (axis cs:2,0) -- (axis cs:2,5) -- (axis cs:0,5);
  \draw[dashed,midgrey] (axis cs:5,0) -- (axis cs:5,2) -- (axis cs:0,2);
  \draw[-{Stealth},navy,thick] (axis cs:2.6,4.6) to[bend left=18] (axis cs:4.6,2.6);
  \node[font=\footnotesize,align=left,anchor=west] at (axis cs:5.4,5.6) {trader buys BTC with USD:\\pool holds less BTC,\\so its price (slope) rises};
  \node[font=\footnotesize,navy] at (axis cs:9.1,1.55) {$xy=k$};
\end{axis}
\end{tikzpicture}
\caption{A constant-product market maker. The pool moves from $P_1$ to $P_2$ along the curve; the quoted price at each point is the slope of the tangent.}
\label{fig:cfmm}
\end{figure}

Decentralized exchanges (DEXs) largely abandoned the order book. Instead, liquidity providers (LPs) deposit reserves $(x,y)$ of two assets into a pool governed by a trading function, and every trade moves the pool along a curve \citep{angeris2020,xu2023sok}:
\begin{equation}
f(x,y)=k, \qquad \text{Uniswap: } x\cdot y = k,\qquad p = -\frac{dy}{dx} = \frac{y}{x}.
\label{eq:cpmm}
\end{equation}
The marginal price is the slope of the curve. Buying the risky asset depletes its reserve and raises its price along a pre-specified path (Figure~\ref{fig:cfmm}). A convex curve approaches both axes asymptotically, so price rises without bound as a reserve becomes scarce, a useful property for an unmanned market maker. \citet{lehar2025} study the equilibrium of Uniswap liquidity provision and its relation to order-book markets.

The economics of the LP resemble those of the classical dealer. Revenue is a fee per trade. Costs are inventory risk and adverse selection: arbitrageurs trade against stale pool prices whenever the external market moves. \citet{milionis2022lvr} formalize this cost as \emph{loss-versus-rebalancing} (LVR), the shortfall of an LP position relative to a continuously rebalanced portfolio; the related ``divergence'' or ``impermanent'' loss measures the LP's shortfall against passive holding. \citet{capponi2021} show that AMM liquidity provision can be unprofitable in volatile, thin markets. Table~\ref{tab:amm} contrasts AMMs with limit order books.

\begin{table}[t]
\centering
\caption{Decentralized exchanges with AMMs versus traditional limit order books. Adapted from \citet{aldridge2025crypto}.}
\label{tab:amm}
\begin{tabularx}{\textwidth}{@{}p{4.2cm}LL@{}}
\toprule
\textbf{Property} & \textbf{DEX with AMM} & \textbf{Traditional exchange} \\
\midrule
Pricing & Constant-function curve & Central limit order book \\
Matching & One-to-pool: traders never meet each other & One-to-one: buyer meets seller \\
Hours & 24$\times$7 & Mostly business hours \\
Quotes vary with & Pool inventory and liquidity & Dealer discretion, uncertainty \\
Price path & Pre-specified by the curve & Depends on market-maker participation \\
Settlement & On-chain, atomic with the trade & T+1 through clearing \\
\bottomrule
\end{tabularx}
\end{table}

\subsection{Batching returns, with a builder in charge}

Ethereum's move to proof of stake in 2022 organized transaction processing into a batch architecture. Pending transactions sit in a public mempool; every 12 seconds, a specialized \emph{block builder} selects transactions and chooses their order; a validator proposes the block with its stake at risk; every trade then fills against an AMM pool (Figure~\ref{fig:pipeline}). The architecture is reminiscent of frequent batch auctions, but with two differences that invert their purpose. First, there is no uniform clearing price: trades within a block execute sequentially along the AMM curve, so their \emph{order} determines their price. Second, the party that sets the order sees the order flow first and may insert its own trades.

\begin{figure}[t]
\centering
\begin{tikzpicture}[font=\footnotesize,
  box/.style={draw=navy,thick,minimum width=2.25cm,minimum height=1.75cm,align=center,text width=2.05cm},
  hl/.style={box,fill=navy,text=white}]
  \node[box] (t) at (0,0) {\textbf{Traders}\\submit orders};
  \node[box,right=0.45cm of t] (m) {\textbf{Mempool}\\pending orders, visible to builders};
  \node[hl,right=0.45cm of m] (b) {\textbf{Block builder}\\selects and orders every 12\,s};
  \node[box,right=0.45cm of b] (v) {\textbf{Validator}\\proposes block; stake at risk};
  \node[box,right=0.45cm of v] (a) {\textbf{AMM pool}\\fills every order};
  \foreach \u/\w in {t/m,m/b,b/v,v/a}{\draw[-{Stealth},navy,thick] (\u) -- (\w);}
  % sandwich
  \node[draw=midgrey,minimum width=3.2cm,minimum height=1.05cm,align=center,text width=3.0cm] (s1) at (0.6,-2.45) {\textbf{1. Builder buys first}\\pool price 100 $\to$ 101};
  \node[draw=navy,fill=lightnavy,minimum width=3.2cm,minimum height=1.05cm,align=center,text width=3.0cm,right=0.9cm of s1] (s2) {\textbf{2. User's buy fills}\\at worse price: 101 $\to$ 103};
  \node[draw=midgrey,minimum width=3.2cm,minimum height=1.05cm,align=center,text width=3.0cm,right=0.9cm of s2] (s3) {\textbf{3. Builder sells}\\pool price 103 $\to$ 102};
  \draw[-{Stealth},thick] (s1) -- (s2); \draw[-{Stealth},thick] (s2) -- (s3);
  \node[darkgrey,anchor=west] at (-1.1,-3.55) {A sandwich attack inside one block (illustrative prices): buy low, sell high around the user, risk-free.};
\end{tikzpicture}
\caption{Top: Ethereum's post-merge transaction pipeline. Bottom: how control of intra-block ordering yields maximal extractable value.}
\label{fig:pipeline}
\end{figure}

The consequence is \emph{maximal extractable value} (MEV): the profit available to whoever controls inclusion and ordering. \citet{daian2020} documented priority-gas-auction bots front-running DEX trades and warned of consensus instability. \citet{qin2022} quantified the extracted value across sandwich attacks, arbitrage and liquidations. The canonical sandwich attack (Figure~\ref{fig:pipeline}, bottom) buys ahead of a pending user order, lets the order execute at a worse price, and sells immediately after. \citet{adams2025mev} decompose the realized cost of Uniswap swaps into benign slippage and adversarial ``reordering slippage''. They find that gas dominates costs for small swaps, while price impact and slippage dominate for large ones. Because the curve's price path is deterministic, a trade and its reversal need not commute, which is the formal basis of reordering attacks \citep{bartoletti2022}.

\citet{aldridge2025crypto} summarizes the irony. Periodic batching was designed to eliminate the need for continuous market makers and the rents of those who move first. On Ethereum it hosts AMMs that must fill every trade and hands first-mover rents to the builder, reproducing the front-running complaints once made about HFT. At the same time, crypto still needs AMMs, because they are what make asynchronous 24/7 trading possible when no counterparty is present.

\subsection{The research response: restrain the builder, or rescue the AMM}

The literature separates two problems: \emph{who orders the batch}, and \emph{who prices the trade}. Table~\ref{tab:defi-lit} organizes representative proposals along this divide.

\begin{table}[t]
\centering
\caption{Representative responses to MEV and AMM unprofitability.}
\label{tab:defi-lit}
\begin{tabularx}{\textwidth}{@{}LL@{}}
\toprule
\textbf{Proposal} & \textbf{Studies} \\
\midrule
\multicolumn{2}{@{}l}{\emph{Restrain the builder: ordering and extraction}} \\
Formal theory of reordering attacks on AMMs & \citet{bartoletti2022} \\
Batch all trades and execute at the post-batch marginal price (function-maximizing AMM), eliminating LVR and sandwiches & \citet{canidio2023} \\
Process each block as a batch under a potential function, making the AMM arbitrage-resilient & \citet{chan2025amm} \\
Dynamic default slippage limits to make sandwiches unprofitable & \citet{chemaya2023} \\
Verifiable ordering that minimizes intra-block price volatility & \citet{mclaughlin2024clvr} \\
\midrule
\multicolumn{2}{@{}l}{\emph{Rescue the AMM: pricing and liquidity provision}} \\
Curves that maximize the number of trades intermediated & \citet{goyal2023} \\
Optimal fees for constant-function market makers & \citet{fritsch2021fees} \\
Mechanism-design framework in which traders report their price expectations & \citet{milionis2024myerson} \\
Predictable loss and optimal liquidity provision via stochastic control & \citet{cartea2024} \\
Missed optimization opportunities in lending, flash swaps and liquidations & \citet{yaish2023} \\
Parallel execution through incentive-compatible sharded pools & \citet{chen2024samm} \\
Programmable ``hooks'' that customize pools on the fly (Uniswap v4) & \citet{bachu2025} \\
\bottomrule
\end{tabularx}
\end{table}

Two strands connect directly to earlier sections. First, several of the most effective MEV remedies are batch auctions with uniform pricing \citep{canidio2023,chan2025amm}: the on-chain literature has rediscovered \citet{budish2015}, now as a defense against the builder rather than the HFT. Second, \citet{milionis2024myerson} bring the revelation principle of Section~\ref{sec:revelation} into AMM design. Each trader reports a price $\hat p$ drawn from a distribution known to the market maker and receives $x(\hat p)=g(p_0)-g(\hat p)$ units of the risky asset for a payment $y(\hat p)$. The designer chooses the demand curve $g^*$ that maximizes expected profit $\mathbb{E}[\,y(\hat p)-p_0\,x(\hat p)\,]$, subject to an update rule $\pi(p_0,\hat p)$ that lies between $p_0$ and $\hat p$, is non-decreasing in $\hat p$, and leaves the price unchanged when $\hat p=p_0$. The result is an order-book-like AMM in which each trader calls her own price, with truthful reporting designed in rather than hoped for.

\subsection{Hiding the book on-chain}

Section~\ref{sec:information} suggests a third direction: rather than restraining whoever sees the order flow, stop publishing it. Table~\ref{tab:onchain-dark} contrasts three designs. In a ``digital dark pool'' layer-1 chain proposed by \citet{aldridge2025crypto}, traders submit prices \emph{and} quantities that are not displayed on-chain, no validator orders transactions, and only execution prices are posted. The design follows the logic of \citet{milionis2024myerson} in that traders call their own prices, but it omits periodic batching, in line with the $W_{\text{DARK}}>W_{\text{LIT}}>W_{\text{BATCH}}$ ranking. Payments-oriented chains with opt-in privacy \citep{circle2025arc} point in a similar direction for settlement. These designs are early, and their equilibrium properties on a permissionless network remain open questions.\footnote{Disclosure: the author is affiliated with AbleBlox, which develops the digital dark pool design described here.}

\begin{table}[t]
\centering
\caption{Three designs compared by what is visible before the trade and who controls ordering.}
\label{tab:onchain-dark}
\begin{tabularx}{\textwidth}{@{}p{3.2cm}LLL@{}}
\toprule
\textbf{Design} & \textbf{Visible before trade} & \textbf{Who orders trades} & \textbf{How trades match} \\
\midrule
Lit exchange & Full order book & Price-time priority & Buyer to seller \\
Ethereum with AMM & Pending orders in the mempool & Block builder, for a fee & Trader to pool \\
Digital dark pool L1 & Nothing; only output prices are posted & A matching rule; no validators & Traders' own prices and quantities, matched directly \\
\bottomrule
\end{tabularx}
\end{table}

% =====================================================================
\section{AI Agents as Market Participants}
\label{sec:agents}
% =====================================================================

\subsection{From algorithms to learning agents}

Algorithmic trading automated the \emph{execution} of human decisions. The current wave automates the decisions themselves, first with reinforcement learning and now with agents built on large language models (LLMs). The leading concern this raises for microstructure is tacit collusion. \citet{calvano2020} showed that Q-learning pricing algorithms in repeated oligopoly learn supra-competitive prices sustained by punishment of deviations, without communication. \citet{dou2025} bring the question to financial markets. Informed speculators trading with reinforcement learning autonomously sustain collusive, supra-competitive profits by under-reacting to information, without agreement, communication or intent, which harms price informativeness and liquidity. They identify two mechanisms. The first is collusion through punishment threats, which arises only when price efficiency and information asymmetry are not too high. The second is collusion through homogenized learning biases (``artificial stupidity''), which persists even in efficient markets. For LLM agents, \citet{agrawal2025} find in simulated continuous double auctions that direct communication between seller agents increases collusive tendencies, that propensities to collude differ across models, and that oversight and urgency instructions change behavior.

These results sharpen the lesson of \citet{gode1993}. Market structure can extract efficiency from traders who are individually naive, but that argument assumes their errors are independent. Learning agents trained on similar data, or running the same underlying model, may share their errors, and a mechanism that is robust to random traders need not be robust to correlated ones.

\subsection{The agent commerce stack}

\begin{figure}[t]
\centering
\begin{tikzpicture}[font=\footnotesize,
  layer/.style={draw=navy,thick,minimum width=8.2cm,minimum height=0.85cm,align=center,text width=7.9cm},
  missing/.style={draw=navy,thick,dashed,minimum width=8.2cm,minimum height=0.85cm,align=center,text width=7.9cm,fill=lightnavy}]
  \node[missing] (alloc) at (0,4.05) {\textbf{Allocation: who gets it when many agents want it?}\\ \emph{no standard; the subject of Section~\ref{sec:matching}}};
  \node[layer] (id) at (0,2.7) {\textbf{Identity and reputation: ERC-8004} (2025)\\is the agent who it claims to be; how has it behaved?};
  \node[layer,fill=navy!8] (pay) at (0,1.35) {\textbf{Payment: x402} (2025)\\can the agent pay for it, over plain HTTP?};
  \node[layer,fill=navy!15] (mcp) at (0,0) {\textbf{Tools and data: MCP} (2024)\\can the agent reach the resource?};
  \node[draw=midgrey,minimum width=8.2cm,minimum height=0.7cm,align=center,text width=7.9cm] (chain) at (0,-1.3) {Settlement layer: blockchains and stablecoins (Section~\ref{sec:onchain})};
  \node[right=0.35cm of alloc,text width=3.6cm,align=left,navy] {open problem};
\end{tikzpicture}
\caption{The emerging protocol stack for agent commerce. Connectivity, payment and identity are standardized; allocation of scarce resources is not.}
\label{fig:stack}
\end{figure}
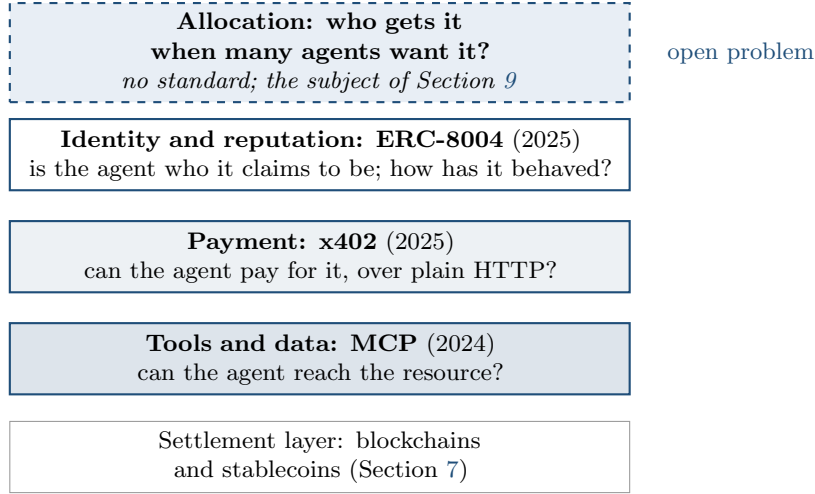

For agents to transact rather than merely chat, three pieces of infrastructure are needed: a way to reach external tools and data, a way to pay without a human in the loop, and a way to be identified and held accountable. Each now has an open standard (Figure~\ref{fig:stack}).

\begin{itemize}[leftmargin=*,itemsep=3pt]
  \item \textbf{Tools: the Model Context Protocol.} Released by Anthropic in November 2024, MCP defines how an agent's client discovers and calls the tools, resources and prompts exposed by any MCP server wrapping a database, file system or API \citep{anthropic2024mcp}. It turns an $M$-agents-by-$N$-tools integration problem into an $M+N$ standardization problem. In December 2025 MCP was donated to the Agentic AI Foundation, a directed fund under the Linux Foundation co-founded by Anthropic, Block and OpenAI \citep{linuxfoundation2025aaif}.
  \item \textbf{Payment: x402.} Launched by Coinbase in May 2025, x402 revives the long-dormant HTTP \texttt{402 Payment Required} status code \citep{coinbase2025x402}. A server answers a request for a paid resource with a 402 response listing payment terms; the client signs a stablecoin payment payload, retries the request with the payload in a header, and a facilitator verifies and settles the payment on-chain before the resource is returned. No account or API key is required.
  \item \textbf{Identity: ERC-8004.} The ``Trustless Agents'' draft standard, proposed in August 2025, defines three on-chain registries \citep{erc8004}. An Identity Registry issues each agent an ERC-721 token that resolves to a registration file. A Reputation Registry records feedback from clients. A Validation Registry lets independent validators record checks of an agent's work.
\end{itemize}

Two observations follow for microstructure. First, the stack settles on the infrastructure of Section~\ref{sec:onchain}. x402 pays in stablecoins and ERC-8004 identities live on-chain, so agent commerce inherits whatever finality, cost predictability and extractable value those chains deliver. Second, the stack is \emph{connective} rather than \emph{allocative}. Table~\ref{tab:stack} makes the gap explicit.

\begin{table}[t]
\centering
\caption{Which question each agent protocol answers.}
\label{tab:stack}
\begin{tabularx}{\textwidth}{@{}LL@{}}
\toprule
\textbf{Question} & \textbf{Answered by} \\
\midrule
Can the agent reach the resource? & MCP \citep{anthropic2024mcp} \\
Can the agent pay for it? & x402 \citep{coinbase2025x402} \\
Is the agent who it claims to be, and has it behaved well? & ERC-8004 \citep{erc8004} \\
Who gets the resource when many agents want the same one? & Not yet answered \\
\bottomrule
\end{tabularx}
\end{table}

\subsection{Evidence from agentic commerce}

Early evidence suggests that leaving allocation to unmediated agent interaction is costly. \citet{allouah2025} evaluate AI shopping agents choosing among products and document systematic biases and strong dependence on the underlying model, with implications for sellers and platforms. \citet{zhu2025agents} model agent-to-agent negotiation and transactions in consumer markets and find that outcomes without a mediating allocation mechanism can be systematically imbalanced across agents. This is exactly the gap that classical market design was built to fill: the allocation question behind double auctions, batch auctions and top trading cycles reappears at the top of the agent stack, now for a non-human population.

\subsection{What agents can cheaply report}

The revelation principle says a designer may restrict attention to direct mechanisms, but a direct mechanism is useful only if agents can report what it asks for. Classical matching mechanisms ask for complete rankings or cardinal utilities over every object. For an agent acting on a user's brief, a ranking over a catalog of hundreds of products is expensive to elicit and to reason about. A short vector of feature weights (price sensitivity, battery life, sound quality, and so on) is cheap. The allocation problem for agents is therefore a problem of \emph{low-dimensional reports}: how to aggregate many feature-weight vectors into one fair and efficient assignment of scarce objects.

% =====================================================================
\section{Allocation Without Rankings: Matching Mechanisms for Agents}
\label{sec:matching}
% =====================================================================

\subsection{Allocation without money}

Some goods cannot, legally or ethically, be allocated by price. \citet{shapley1974} posed the housing market, in which each agent owns one house and may prefer another's. Gale's Top Trading Cycles (TTC) algorithm has each agent point to the owner of her most preferred remaining house, trades along every resulting cycle, removes those agents, and repeats. TTC is Pareto efficient, individually rational and strategy-proof, and \citet{ma1994} proved it is the \emph{unique} mechanism with all three properties. The same cycle logic underlies school choice \citep{abdulkadiroglu2003} and kidney exchange, where incompatible patient--donor pairs swap donors along cycles, and chains started by altruistic donors clear the pool further \citep{roth2004,roth2005}. Deferred acceptance \citep{gale1962} plays the corresponding role for two-sided markets. TTC shows that removing money does not require giving up efficiency or truthful incentives.

With cardinal utilities the picture changes. \citet{hylland1979} allocate probability shares through a pseudo-market with budgets and prices, obtaining ex-ante efficiency but not strategy-proofness. \citet{zhou1990} proved that no mechanism for cardinal one-sided matching is simultaneously ex-ante efficient, strategy-proof and symmetric. Subsequent work seeks approximate versions of all three, usually measured by Nash social welfare (NSW), the product of agents' gains, which balances efficiency and fairness \citep{nash1950,caragiannis2019}. Table~\ref{tab:matching} summarizes representative mechanisms. All of the cardinal ones require agents to report complete utilities over objects.

\begin{table}[t]
\centering
\caption{Cardinal matching mechanisms and what they require agents to report. Adapted from \citet{aldridge2026matching}.}
\label{tab:matching}
\small
\begin{tabularx}{\textwidth}{@{}p{3.1cm}p{2.3cm}LLL@{}}
\toprule
\textbf{Mechanism} & \textbf{Agents report} & \textbf{Method} & \textbf{Welfare guarantee} & \textbf{Incentives} \\
\midrule
\citet{hylland1979} & Utilities over objects & Pseudo-market with budgets and prices & Ex-ante Pareto optimal & Strategic bidding possible \\
\citet{chawla2010} & Multi-parameter valuations & Sequential posted prices, one dimension at a time & $O(1)$ approximation & Truthful in expectation \\
\citet{devanur2015} & Utilities over objects & Random assignment within a configuration LP & 2-approximation to NSW & Envy-free in expectation \\
\citet{abebe2020} & Utilities over objects & Random sampling and configuration LPs & $O(1)$ approximation to NSW & Truthful in expectation \\
Spectral feature matching \citep{aldridge2026matching} & Feature weights & One SVD projection, then a sort & Exact projected-NSW optimum; conditional true-NSW bound & Stable to noise; not strategy-proof \\
\bottomrule
\end{tabularx}
\end{table}

\subsection{A feature-based model}

Consider $I$ agents and $J$ objects with capacities $M_j$ summing to $I$. Objects are described by feature vectors $f_j\in\mathbb{R}^X$ and agents by preference weights $u_i\in\mathbb{R}^X$, with $X$ small. Utility is the inner product
\begin{equation}
U_{ij} \;=\; u_i\cdot f_j \;=\; \sum_{x=1}^{X} u_{ix} f_{jx},
\end{equation}
which is additive across features and so rules out complementarities. An agent reports $w_i$, which may differ from $u_i$. Fairness is measured relative to uniform random assignment. Agent $i$'s disagreement point is $o_i=\frac1J\sum_j U_{ij}$, and
\begin{equation}
\mathrm{NSW}(P\mid U) \;=\; \prod_{i=1}^{I}\big(\mathbb{E}[U_i\mid P]-o_i\big),
\end{equation}
defined to be zero if any agent falls below her disagreement point. An NSW maximizer is Pareto efficient, leaves no agent worse off than random assignment, and is invariant to positive rescaling of any agent's utilities. For reporting, \citet{aldridge2026matching} fixes a clipped version, $\mathrm{NSW}_\epsilon=\prod_i\max(\mathrm{gain}_i,\epsilon)$ with $\epsilon=0.01$, which is always disclosed alongside the rate of individual-rationality (IR) violations. This matters because clipping makes $\mathrm{NSW}_\epsilon$ optimistic precisely when agents are being harmed.

\subsection{Spectral projection}

Collect object features in $F\in\mathbb{R}^{J\times X}$ with singular value decomposition $F=U\Sigma V^\top$. By the Eckart--Young theorem \citep{eckart1936}, the leading right singular vector $v_1$ is the direction of maximum variance of the object features, and the rank-one approximation $\sigma_1u_1v_1^\top$ minimizes reconstruction error. The mechanism projects both sides onto $v_1$ and sorts.

\begin{algorithm}[t]
\caption{Spectral feature matching \citep{aldridge2026matching}}
\label{alg:svd}
\begin{algorithmic}[1]
\Require feature matrix $F\in\mathbb{R}^{J\times X}$; reports $w_1,\dots,w_I\in\mathbb{R}^X$; capacities $M_j$
\State Compute $v_1$, the leading right singular vector of $F$ (oriented so that its entries are predominantly positive)
\State Gauge-fix reports: $\hat w_i \gets w_i/\lVert w_i\rVert$ \Comment{removes dependence on report scale}
\State Project: $\tilde f_j \gets f_j\cdot v_1$ for all $j$; \ $\hat a_i \gets \hat w_i\cdot v_1$ for all $i$
\State \textbf{Round diagnostic:} $n_+ \gets \#\{i:\hat a_i>0\}$; \ $C_+ \gets \sum_{j:\tilde f_j>\bar f} M_j$ where $\bar f$ is the mean projected score
\If{$n_+\neq C_+$} \Return \textsc{Fail} \Comment{no deterministic allocation can be individually rational for all}
\EndIf
\State Within each sign class, sort agents by $\hat a_i$ and objects by $\tilde f_j$ and match in order, respecting capacities
\State \textbf{Margin diagnostic:} report $\mu^*$, the minimum projected gain, against $4\Delta$ with $\Delta=\sqrt{X-1}\,\sigma_2$
\end{algorithmic}
\end{algorithm}

The procedure runs in $O(\min(J^2X,JX^2)+I\log I+J\log J)$ time, dominated by the SVD and two sorts. It is deterministic and symmetric by construction: identical reports receive identical treatment.

\paragraph{The gauge.} A naive version that projects raw reports has a scale bug. Because $\hat a_i$ scales with $\lVert w_i\rVert$, rescaling one agent's report without changing her preferences can flip the entire allocation and move NSW from positive to zero. Normalizing reports to unit length before projecting (a ``gauge'' choice) removes the dependence and restores the scale invariance that NSW itself enjoys.

\begin{figure}[t]
\centering
\includegraphics[width=0.62\textwidth]{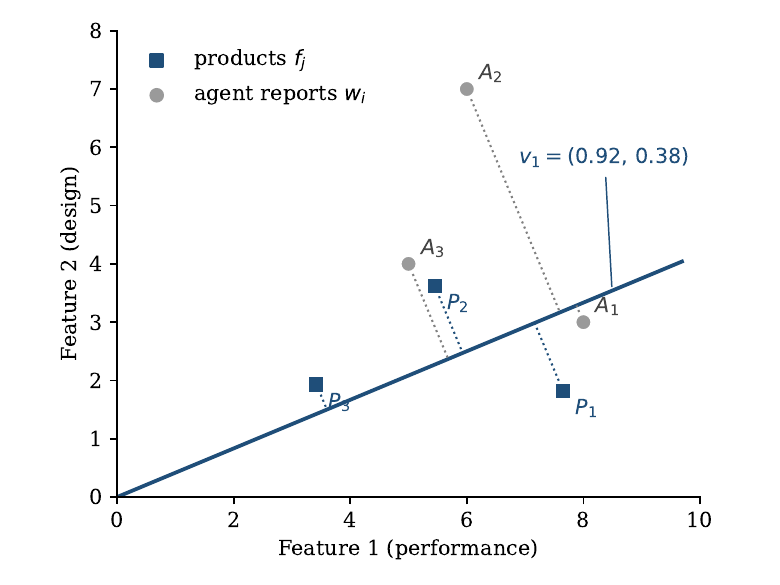}
\caption{Worked example: products (squares) and agent reports (circles) projected onto the leading singular vector $v_1$ of the product-feature matrix. Values recomputed from the example in \citet{aldridge2026matching}.}
\label{fig:toy}
\end{figure}

\paragraph{A worked example.} Take three products with features (performance, design) $f_1=(7.65,1.82)$, $f_2=(5.45,3.62)$, $f_3=(3.42,1.93)$, and three agents reporting $w_1=(8,3)$, $w_2=(6,7)$, $w_3=(5,4)$ \citep[Example~8.1]{aldridge2026matching}. The singular values are $\sigma_1=10.80$ and $\sigma_2=1.83$, and $v_1=(0.92,0.38)$: products differ mainly in performance (Figure~\ref{fig:toy}). Projected product scores are $7.76$, $6.42$ and $3.90$. Raw projected agent scores are $8.54$, $8.23$ and $6.15$, so plain sorting assigns $A_1\to P_1$, $A_2\to P_2$, $A_3\to P_3$. After the gauge fix the scores become $0.999$, $0.893$ and $0.961$, and agents~2 and~3 swap places, which illustrates how much report scale matters.

The example also illustrates the mechanism's central negative result. All three agents rank the products identically ($P_1\succ P_2\succ P_3$), so every allocation must hand the weakest product to someone. Table~\ref{tab:toy} enumerates all six assignments: none leaves every agent above her disagreement point, and every assignment has NSW equal to zero. The round diagnostic detects this in advance. All three gauge-fixed scores are positive ($n_+=3$), but only two products lie above the mean projected score ($C_+=2$), so the algorithm returns \textsc{Fail}.

\begin{table}[t]
\centering
\caption{All deterministic assignments in the worked example. Gains are measured against each agent's random-assignment disagreement point $o=(51.42,\,50.24,\,37.36)$. No assignment is individually rational for all three agents. The plain sorted assignment (first row) maximizes total utility.}
\label{tab:toy}
\begin{tabular}{@{}lrrrr@{}}
\toprule
\textbf{Assignment} ($A_1,A_2,A_3$) & \textbf{Gain $A_1$} & \textbf{Gain $A_2$} & \textbf{Gain $A_3$} & \textbf{Total utility} \\
\midrule
$P_1,P_2,P_3$ & 15.24 & 7.80 & $-12.54$ & 149.52 \\
$P_1,P_3,P_2$ & 15.24 & $-16.21$ & 4.37 & 142.42 \\
$P_2,P_1,P_3$ & 3.04 & 8.40 & $-12.54$ & 137.92 \\
$P_2,P_3,P_1$ & 3.04 & $-16.21$ & 8.17 & 134.02 \\
$P_3,P_1,P_2$ & $-18.27$ & 8.40 & 4.37 & 133.52 \\
$P_3,P_2,P_1$ & $-18.27$ & 7.80 & 8.17 & 136.72 \\
\bottomrule
\end{tabular}
\end{table}

\subsection{Guarantees and diagnostics}

\citet{aldridge2026matching} establishes the following properties for the gauge-fixed mechanism. We state them informally and refer to the paper for precise conditions.
\begin{enumerate}[leftmargin=*,itemsep=3pt]
  \item \textbf{Exact projected optimum.} Within the projected one-dimensional space, sorted matching by sign class computes the exact NSW-maximizing allocation. The result is an equality, not an approximation.
  \item \textbf{Unconditional utilitarian guarantee.} Total true utility is bounded relative to the utilitarian optimum without any individual-rationality claim.
  \item \textbf{Conditional NSW guarantee.} A multiplicative bound on true NSW holds when the feasibility condition $n_+=C_+$ is met and the minimum projected gain clears a margin proportional to the projection error, $\mu^*\ge 4\Delta$ with $\Delta=\sqrt{X-1}\,\sigma_2$.
  \item \textbf{Necessity.} When $n_+\neq C_+$, \emph{no} deterministic mechanism can guarantee positive NSW. Agents who agree too strongly about which objects are best leave no room for an allocation that beats random assignment for everyone. This is the same structural obstacle, a shared ranking, that drives the impossibility results of Section~\ref{sec:foundations}.
  \item \textbf{Incentives.} The mechanism is stable to exogenous reporting noise but is \emph{not} strategy-proof, and the paper gives an explicit profitable misreport.
\end{enumerate}

Two diagnostics make the conditional guarantee usable. At the \emph{market} level, the share of variance on the leading direction, $\rho_1=\sigma_1^2/\sum_\ell\sigma_\ell^2$, indicates whether ``generically better'' is a meaningful axis; the paper uses $\rho_1\ge0.5$ as a deployment threshold. At the \emph{round} level, the feasibility test $n_+=C_+$ and the margin test $\mu^*\ge4\Delta$ are computed from the algorithm's own output before any true utilities are observed. Contrast Myerson--Satterthwaite: there the efficiency loss is unconditional and unmeasured. Here it is conditional, and the failure regime is identified exactly and can be checked in advance.

\subsection{Evidence from a simulated agentic-shopping market}

\citet{aldridge2026matching} tests the mechanism on the application motivating this section: LLM-style shopping agents, each translating a user's brief into feature weights over six headphone attributes, compete for a limited drop of ten products, one unit each.

\paragraph{A single drop.} The full 24-product catalog looks comfortably low-rank ($\rho_1=0.912$, effective rank $2.24$, with $v_1$ loading positively on all six features). The ten-product drop still passes the market-level threshold ($\rho_1=0.871$), but its projection-error bound $\Delta=35.2$ is large relative to individual agents' gains. This is something the market-level $\rho_1$ cannot reveal. With only mainstream shoppers (Scenario~A), all ten gauge-fixed scores are positive but only six products lie above the mean ($n_+=10\neq C_+=6$), and the mechanism correctly returns \textsc{Fail}. With four contrarian shoppers substituted (Scenario~B), feasibility holds ($n_+=C_+=6$) and total utility reaches 99.1\% of the utilitarian optimum (1{,}467.6 versus 1{,}480.8). The margin diagnostic nevertheless fails by three orders of magnitude ($\mu^*=0.26$ against $4\Delta=140.8$). It correctly predicts that one agent ends below random assignment and that the mechanism's $\mathrm{NSW}_\epsilon$ reaches only about 0.27\% of the true optimum found by brute force over all $10!\approx3.6$ million assignments. A blind search over 200{,}000 alternative reports finds a misreport that improves the harmed agent's true gain by 33.6. Truthful reporting is not a best response.

\begin{figure}[t]
\centering
\begin{minipage}[b]{0.49\textwidth}
\centering
\includegraphics[width=\linewidth]{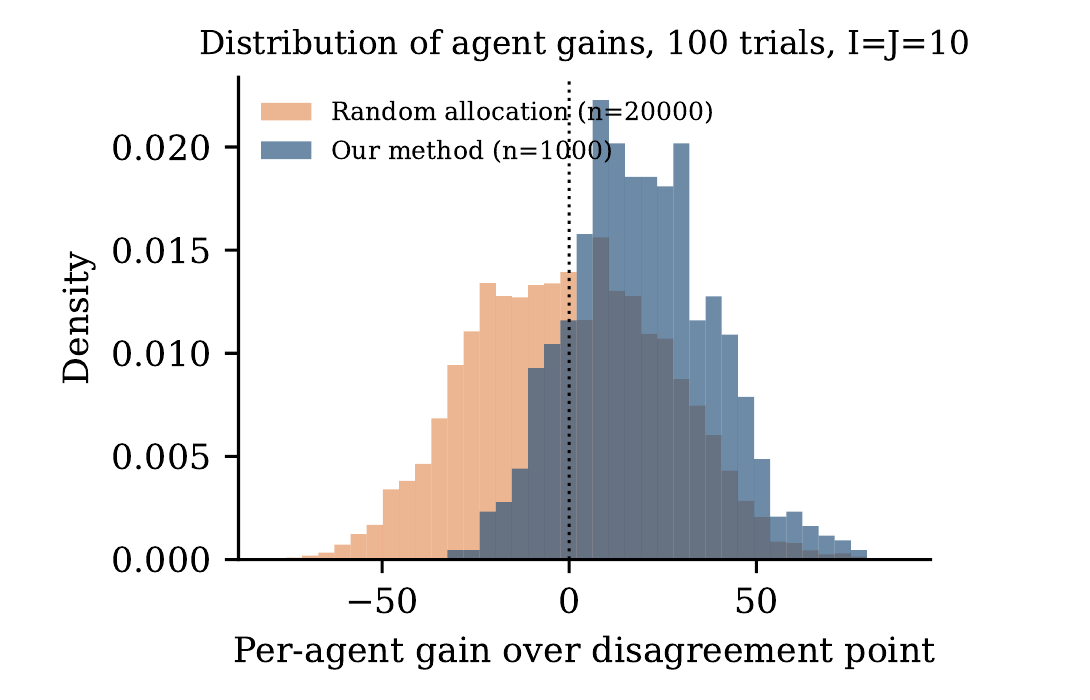}
\captionof{figure}{Per-agent gains over the disagreement point, pooled over 100 random markets; the mechanism (blue) versus random allocation (orange). Source: \citet{aldridge2026matching}, Figure~3.}
\label{fig:gains}
\end{minipage}\hfill
\begin{minipage}[b]{0.48\textwidth}
\centering
\begin{tikzpicture}
\begin{axis}[ybar, bar width=15pt, width=\linewidth, height=5.3cm,
  ymin=0, ymax=25, ylabel={Mean $\log \mathrm{NSW}_\epsilon$},
  symbolic x coords={Random,Serial dict.,Spectral,Utilitarian opt.},
  xtick=data, x tick label style={font=\scriptsize,align=center,text width=1.3cm},
  tick label style={font=\scriptsize}, label style={font=\footnotesize},
  nodes near coords, every node near coord/.append style={font=\scriptsize,yshift=6pt},
  axis x line*=bottom, axis y line*=left, enlarge x limits=0.18, ymajorgrids, grid style={gray!20}]
\addplot[fill=navy,draw=navy,error bars/.cd,y dir=both,y explicit]
  coordinates {(Random,6.52) +- (0,1.23) (Serial dict.,17.95) +- (0,1.72) (Spectral,17.37) +- (0,1.70) (Utilitarian opt.,20.01) +- (0,1.67)};
\end{axis}
\end{tikzpicture}
\captionof{figure}{Mean $\log\mathrm{NSW}_\epsilon$ ($\epsilon=0.01$) with 95\% confidence intervals over 100 markets. Mean IR violations per 10 agents: 5.03, 1.48, 1.56 and 1.28. Source: \citet{aldridge2026matching}, Table~6.}
\label{fig:nsw}
\end{minipage}
\end{figure}

\paragraph{One hundred random markets.} Across 100 independently generated markets, the exact feasibility test passes in 22 trials. This is a knife-edge integer condition, neither routine nor rare. Pooled per-agent gains shift clearly rightward relative to random allocation (Figure~\ref{fig:gains}): the mean gain is 19.44 versus $-0.02$ under random assignment (zero by construction), and the share of agents below their disagreement point falls from 50.3\% to 15.6\%. On mean $\log\mathrm{NSW}_\epsilon$ (Figure~\ref{fig:nsw}) the mechanism beats random assignment by about 11 log-points and is statistically indistinguishable from serial dictatorship. Serial dictatorship uses true, unprojected utilities and requires one best response per agent, but carries no formal guarantee.

\paragraph{Deployment rules.} The evidence supports four rules for agentic commerce. First, diagnose every round, not just the market, because a small, scarce drop can be far worse conditioned than the catalog it comes from. Second, run both round checks before allocating. Third, fall back to exact search when they fail; for a ten-unit drop, brute force took 16.5 seconds, affordable for small, high-stakes rounds. Fourth, do not assume that agents report truthfully: production systems need an incentive layer or explicit disclosure to users.

% =====================================================================
\section{Synthesis: What Each Design Buys and What It Costs}
\label{sec:synthesis}
% =====================================================================

Table~\ref{tab:synthesis} lines up the designs surveyed. The throughline is that every gain in speed, scale or simplicity is paid for somewhere else in the mechanism.

\begin{table}[t]
\centering
\caption{Eight designs, one throughline: every gain is paid for elsewhere in the mechanism.}
\label{tab:synthesis}
\begin{tabularx}{\textwidth}{@{}p{3.3cm}LL@{}}
\toprule
\textbf{Design} & \textbf{What it buys} & \textbf{What it costs} \\
\midrule
Double auction & Truthfulness or budget balance & Exact efficiency \citep{myerson1983} \\
Frequent batch auction & No latency arms race & Timely price discovery; waiting costs; often not price-forming in practice \\
Dark pool & No book-timing games; highest welfare at moderate arrival rates & Participation of uninformed marginal traders \\
Bilateral / RFQ market & Trading in illiquid, customized instruments; discretion & Search costs, dealer bargaining power, no continuous public price \\
LMSR / AMM & Liquidity with thin or asynchronous trading & A bounded subsidy; loss-versus-rebalancing for LPs; displaced by order books once volume arrives \\
On-chain AMM with builder & 24$\times$7 atomic settlement & MEV for whoever orders the batch \\
Top trading cycles & Efficiency and truthfulness without money & Requires full ordinal rankings; no transfers \\
Feature matching & Speed and cheap elicitation for agents & A conditional, disclosed guarantee; manipulable \\
\bottomrule
\end{tabularx}
\end{table}

Five themes recur across the eras.

\begin{enumerate}[leftmargin=*,itemsep=4pt]
  \item \textbf{Impossibility relocates rather than disappears.} The theorems of Myerson and Satterthwaite, of Gibbard and Satterthwaite, and of Zhou recur in new clothing. The designer's real choice is where the loss falls and whether it is visible.
  \item \textbf{The binding constraint moves with technology.} For human floor traders the constraint was rationality and information, and \citet{gode1993} showed that structure compensates for the first. For HFT it was speed. On blockchains it is control over ordering. For AI agents it is the dimensionality of what can be reported.
  \item \textbf{Whoever controls the queue captures the rent.} The first arrival in a continuous book, the builder of an Ethereum block and the agent that negotiates without a mediator all profit from sequencing. Batch auctions, opaque books and sort-based matching are different ways of neutralizing sequencing power.
  \item \textbf{Information structure is a design lever, not a given.} Hiding the book raises welfare in equity markets \citep{aldridge2026mechanisms}, and publishing the mempool creates MEV on-chain. The same lever should be available to designers of agent marketplaces.
  \item \textbf{Diagnostics complement guarantees.} When exact guarantees are impossible, the next best thing is a mechanism that says, before the outcome, whether its guarantee holds. The round-level diagnostics of Section~\ref{sec:matching} are one instance; pre-trade measures of MEV exposure or of participation risk in dark venues would be others.
\end{enumerate}

% =====================================================================
\section{Conclusion}
\label{sec:conclusion}
% =====================================================================

Market microstructure has always been a discipline of trade-offs imposed by impossibility theorems and relaxed by clever design. The age of AI does not change that logic, but it changes its inputs. Participants are faster, more numerous and more correlated. They report less than full preferences. They transact over protocols that standardize connectivity, payment and identity while leaving allocation unaddressed. The designs surveyed here, from Vickrey's second-price rule to spectral feature matching, show that markets can be made to work under each new constraint, provided the designer is explicit about which property is being given up and makes the resulting loss visible. For markets populated by AI agents, that means choosing information structures and sequencing rules deliberately, preferring mechanisms that announce in advance when their guarantees fail, and treating allocation as a first-class layer of agent infrastructure.

\bibliographystyle{plainnat}
\bibliography{references}

\end{document}